\documentclass[a4paper,12pt]{article}
\usepackage[utf8]{inputenc}
\usepackage{amsmath}
\usepackage{amssymb}
\usepackage{mathrsfs}
\usepackage{chemarrow}
\usepackage{times}
\usepackage{graphicx}
\begin{document}

\def\mA{{\bf A}}
\def\vV{{\bf V}}
\def\MS{{\bf S}}
\def\rd{{\rm d}}
\def\mF{{\bf F}}
\def\mQ{{\bf Q}}
\def\vJ{{\bf J}}
\def\rvB{{\bf B}}
\def\bplus{{\bf +}}
\def\bplus{\mbox{\boldmath$+$}}
\def\ve{{\bf e}}
\def\pp{{\bf p}}
\def\vl{{\bf l}}
\def\vn{{\bf n}}
\def\vv{{\bf v}}
\def\vx{{\bf x}}
\def\vz{{\bf z}}
\def\vnu{\mbox{\boldmath$\nu$}}
\def\vxi{\mbox{\boldmath$\xi$}}
\def\vkappa{\mbox{\boldmath$\kappa$}}


\title{Nonequilibrium Thermodynamic Formalism of
Nonlinear Chemical Reaction Systems with
Waage-Guldberg's Law of Mass Action
}

\author{Hao Ge\footnote{haoge@pku.edu.cn}\\[6pt]
Beijing International Center for Mathematical Research (BICMR)\\
and Biodynamic Optical Imaging Center (BIOPIC)\\
Peking University, Beijing 100871, P.R.C.\\[10pt]
and\\[10pt]
Hong Qian\footnote{hqian@u.washington.edu}\\[6pt]
Department of Applied Mathematics\\
University of Washington, Seattle\\
WA 98195-3925, U.S.A\\[10pt]
}

\maketitle

\abstract{Macroscopic entropy production $\sigma^{(tot)}$
in the general nonlinear isothermal chemical reaction system
with mass action kinetics is decomposed into a free energy
dissipation and a house-keeping heat: $\sigma^{(tot)}=\sigma^{(fd)}+\sigma^{(hk)}$;
$\sigma^{(fd)}=-\rd A/\rd t$, where $A$ is a generalized
free energy function. This yields a novel nonequilibrium free
energy balance equation
$\rd A/\rd t=-\sigma^{(tot)}+\sigma^{(hk)}$, which is on a
par with celebrated entropy balance equation
$\rd S/\rd t=\sigma^{(tot)}+\eta^{(ex)}$ where $\eta^{(ex)}$ is
the rate of entropy exchange with the environment.
For kinetic systems with complex
balance, $\sigma^{(fd)}$ and $\sigma^{(hk)}$ are the
macroscopic limits of stochastic free energy dissipation
and house-keeping heat, which are both nonnegative, in the Delbr\"uck-Gillespie description of the stochastic chemical kinetics.
Therefore, we show that a full kinetic and thermodynamic theory of chemical
reaction systems that transcends mesoscopic and macroscopic
levels emerges.}



\section{Introduction}

	Inspired by the recent discovery of three non-negative
entropy productions in mesoscopic, stochastic nonequilibrium
thermodynamics, $e_p=f_d+Q_{hk}$, interpreted as an
equation of free energy balance: $\rd F^{(meso)}/\rd t
\equiv$ $-f_d=Q_{hk}-e_p$, where $e_p$, $Q_{hk}$,
$F^{(meso)}$, and $f_d$ are called entropy production,
house-keeping heat, mesoscopic free energy and free energy
dissipation \cite{esposito2007,ge-09,ge-qian-pre10,esposito-vandenbroeck,qian_jmp,ge-qian-pre13},
we consider the formal kinetics of a general chemical
reaction system
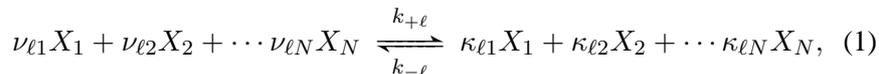
\begin{equation}
     \nu_{\ell 1}X_1+\nu_{\ell 2}X_2 + \cdots
     \nu_{\ell N}X_N  \  \  \underset{k_{-\ell}}{\overset{k_{+\ell}}{\rightleftharpoonsfill{26pt}}}   \  \
      \kappa_{\ell 1}X_1+\kappa_{\ell 2}X_2 + \cdots
     \kappa_{\ell N}X_N,
\label{rxn}
\end{equation}
in which $1\le\ell\le M$: There are $N$ species and
$M$ reactions. $(\kappa_{ij}-\nu_{ij})$ are
{\em stoichiometric coefficients} that relate species to
reactions.  According to Waage-Guldberg's Law of Mass
Action for a macroscopic reaction vessel, at a constant
temperature, with rapidly stirred chemical
solutions, the concentrations of the species at time $t$,
$x_i(t)$ for $X_i$, satisfy the system of ordinary differential
equations \cite{waage-guldberg}
\begin{equation}
  \frac{\rd x_i(t)}{\rd t} = \sum_{\ell=1}^M
               \Big(\kappa_{\ell i}-\nu_{\ell i}\Big)
           \Big(J_{+\ell}(\vx)-J_{-\ell}(\vx)\Big),
\label{the-lma}
\end{equation}
with $\vx=(x_1,x_2,\cdots,x_N)$ and
\begin{equation}
     J_{+\ell}(\vx) = k_{+\ell}\prod_{i=1}^N x_i^{\nu_{\ell i}}, \  \
     J_{-\ell}(\vx) = k_{-\ell}\prod_{i=1}^N x_i^{\kappa_{\ell i}}.
\end{equation}
For a meaningful thermodynamic analysis, we shall assume in
the present paper that $k_{+\ell}=0$ if and only if $k_{-\ell}=0$.

The kinetics of such a chemical reaction system can be very complex.
The simple and well-understood cases are linear, unimolecular reaction
systems, or nonlinear systems whose steady states are detail
balanced  \cite{gnlewis}.  For the latter, it can be shown that
the steady state is unique and the net flux in each and every
reversible reaction is zero \cite{shear-jtb,shear-jcp}.  Therefore, it
is an equilibrium steady state.  Furthermore,
the existence of a chemical equilibrium with detailed balance dictates that
the rate constants $\{k_{\pm\ell}\}$ in such a system satisfy the
Wegscheider-Lewis cycle condition \cite{shapiro-1965,schuster-schuster,qian-beard-liang}.

	J. W. Gibbs was the first to formulate a {\em free energy function}
and showed that Waage-Guldberg's mass action law was
closely related to a variational principle with respect to that function,
connecting thermodynamics with kinetics \cite{shapiro-1965}.
In units of $k_BT$ and per unit volume, the Gibbs function for
a dilute solution \cite{Fermi}:
\begin{equation}
   G\big[\vx\big] = \sum_{j=1}^N x_j\Big(\mu_j - 1\Big), \
        \  \mu_j =
        \mu_j^o+\ln x_j,
\label{eq11}
\end{equation}
in which $\mu_j^o$ is a constant associated with the structure
of the $j^{th}$ chemical species in aqueous solution,
and the $-1$ term is the contribution
from the solvent.  See Appendix A for more discussions.

If all the reaction rate constants $k_{\pm\ell}$ satisfy
Wegscheider-Lewis cycle condition, the
chemical potential {\em difference} \cite{qian-beard-05}:
\begin{equation}
     \Delta\mu_{\ell}\big[\vx^{eq}\big]
        \equiv \sum_{j=1}^N \big(\kappa_{\ell j}-\nu_{\ell j}\big)
            \left(\mu_j^o + \ln x_j^{eq} \right)
        = 0,
\end{equation}
in which $\{x_j^{eq}\}$ is the equilibrium concentration. Then for the
reaction system, with a constant volume, that is away from its
equilibrium at time $t$,
\begin{eqnarray}
  \frac{\rd}{\rd t}G\big[\vx(t)\big]
       &=& \sum_{j=1}^N \frac{\rd x_j(t)}{\rd t}
                      \Big(\mu_j^o + \ln x_j \Big)
\label{G4solution}\\
	&=&  \sum_{j=1}^N\sum_{\ell=1}^M
                     \big( \kappa_{\ell j}
                 -\nu_{\ell j}\big)
               \Big(J_{+\ell}(\vx)-J_{-\ell}(\vx) \Big)
               \Big(\mu_j^o + \ln x_j \Big)	
\nonumber\\
	&=&  \sum_{\ell=1}^M
               \Big(J_{+\ell}(\vx)-J_{-\ell}(\vx) \Big)
              \ln\left( \frac{J_{-\ell}(\vx) }{
                     J_{+\ell}(\vx)}\right) \ \le \ 0.
\label{lyapunov11}
\end{eqnarray}

For open chemical systems that do not reach an equilibrium
with detailed balance, Horn and Jackson introduced
the notion of {\em complex balanced} reaction network
in 1972 \cite{horn-jackson}.  It is a generalization
of both linear reaction networks and kinetics with
detailed balance \cite{horn-1972,feinberg-1972}.
Complex balanced kinetics can be nonlinear as well
as having nonequilibrium steady states (NESS).
It also has a deep relation to the topological structure
of a reaction network \cite{feinberg-91,polettini-esposito-2}.
A complex balanced reaction system has a unique positive
steady state.

For nonequilibrium chemical thermodynamics, how,
or whether even possible, to generalize Gibbs'
approach, in the framework of the mass-action kinetics, to
nonlinear kinetic systems without detailed balance
has remained elusive.  Such systems include the important
class of NESS which is aptly applicable to cellular biochemistry
in homeostasis \cite{qian-beard-05}. L. Onsager's phenomenological
theory is only applicable to systems in the linear regime near an
equilibrium \cite{onsager}; T. L. Hill's NESS thermodynamics
\cite{tlhill} and the grand canonical approach developed in
\cite{heuett-qian} are applicable only to macroscopic linear
chemical kinetics.   But thanks to the recent development in
both mesoscopic, stochastic nonequilibrium
thermodynamics and the resurgent interests in the
stochastic description of nonlinear mass-action kinetic
systems, a cross-fertilization is possible.

In this paper, we revisit the notion of macroscopic, chemical
reaction entropy production $\sigma^{(tot)}[\vx]$
\cite{qian-beard-05,ross-book,degroot-mazur-book},
and show it can also be decomposed into two
parts $\sigma^{(tot)}[\vx]=\sigma^{(fd)}[\vx]+
\sigma^{(hk)}[\vx]$, in which
$\sigma^{(fd)}[\vx]$ is the negative time derivative
of a generalized free energy function $A[\vx]$.  More
interestingly, both $\sigma^{(fd)}[\vx]$ and $\sigma^{(hk)}[\vx]$ can
be mathematically proven as non-negativity for kinetic systems with
complex balance.  Since the $A[\vx]$ is defined
with respect to a positive steady state of  the kinetic
system,  it is no longer unique for systems with multi-stability.
In fact, for a system with multi-stability, one can define $A[\vx]$ with respect to one of the stable steady states, then it
necessarily has negative $\sigma^{(fd)}[\vx]$
for some $\vx$, thus $\sigma^{(hk)}>\sigma^{(tot)}$
at the $\vx$.
We further show that for complex balanced kinetic
systems these natually defined macroscopic
quantities are the macroscopic limits of
the mesoscopic free energy dissipation $f_d$ and house-keeping
heat $Q_{hk}$, according to the stochastic kinetic description of
the same chemical kinetics.

\section{Nonequilibrium thermodynamics of
chemical reaction network}

In the present paper, we do not assume the rate constants $k_{\pm\ell}$ satisfy Wegscheider-Lewis cycle condition unless
stated otherwise.  We do assume, however, that the macroscopic
kinetic system (\ref{the-lma}) has a positive steady state
$\vx^{ss}=\{x^{ss}_i, 1\le i\le N\}$.  Motivated by the recent studies
on mesoscopic, stochastic thermodynamics \cite{esposito2007,ge-qian-pre10,qkkb},
we introduce a decomposition of the instantaneous rate of total
entropy production of the mass-action kinetic system following
Eq. \ref{the-lma}, $\sigma^{(tot)}[\vx]$
\cite{beard-qian-PLoS1,ross-book,degroot-mazur-book},
into two nonequilibrium components, a {\em house-keeping
heat} part \cite{oono-paniconi,HS,Maes2015} and
a {\em free energy dissipation} part:
\begin{subequations}
\label{3q}
\begin{eqnarray}
       \sigma^{(tot)}\big[\vx\big] &=& \sum_{\ell=1}^M
             \Big(J_{+\ell}(\vx)-J_{-\ell}(\vx)\Big)
            \ln \left(\frac{J_{+\ell}(\vx)}{J_{-\ell}(\vx)} \right)
        \ =  \ \sigma^{(hk)}+\sigma^{(fd)},
\label{sigmatot}
\\
	   \sigma^{(hk)}\big[\vx\big] &=&
           \sum_{\ell=1}^M
             \Big(J_{+\ell}(\vx)-J_{-\ell}(\vx)\Big) \ln\left(
           \frac{J_{+\ell}\big(\vx^{ss}\big)}{J_{-\ell}\big(\vx^{ss}\big)}\right),
\\
	 \sigma^{(fd)}\big[\vx\big] &=&  	
        \sum_{\ell=1}^M
                 \Big(J_{+\ell}(\vx)-J_{-\ell}(\vx)\Big)
                   \ln \left(\frac{J_{+\ell}(\vx)J_{-\ell}(\vx^{ss})}{J_{-\ell}(\vx)J_{+\ell}(\vx^{ss})} \right).
\label{sigmafd}
\end{eqnarray}
\end{subequations}

	Both $\sigma^{(tot)}\big[\vx\big]$ and $\sigma^{(hk)}\big[\vx\big]$
have the generic form of ``net flux in reaction $\ell$'' $\times$
``thermodynamic force per $\ell^{th}$ reaction''.  The
latter are expressed in terms of the ratio of
forward and backward one-way fluxes.  This is an insight that
goes back at least to T. L. Hill \cite{tlhill} if not earlier, as
discussed in \cite{beard-qian-PLoS1}.  For the term in
(\ref{3q}b), we adopt the idea of Hatano and Sasa
who first introduced housekeeping heat as the product of
transient thermodynamic fluxes and steady-state thermodynamic
force \cite{HS}.  Since then, there are several different definitions
under the same name \cite{Sasa08,Sasa2015}.  We also note that
the one-way fluxes in chemical kinetics are nonlinear functions
of concentrations in general, while one-way-fluxes in mesocopic
stochastic dynamics are always linear functions of state probabilities,
similar to a unimolecular reaction network.  Thermodynamic forces
of a wide range of processes have a unifying expression in
terms of one-way-fluxes \cite{beard-qian-PLoS1}.

We also note that all three quantities in (\ref{3q})
can be explicitly computed if the
rate laws $J_{\pm\ell}(\vx)$ as well as a kinetic steady state $\vx^{ss}$
are known. Before introducing a further assumption of complex
balanced kinetics in Sec. \ref{sec:compba}, we first
discuss key characteristics of the three macroscopic
quantities introduced in Eq. \ref{3q}.

\subsection{Key characteristics of the three macroscopic
quantities}
\label{sec:2.1}

First, the foremost, $\sigma^{(tot)}[\vx]\ge 0$. It is zero if and
only if at an $\vx$, $J_{+\ell}(\vx)=J_{-\ell}(\vx)$ $\forall\ell$.
This implies the $\vx$ is a steady state of (\ref{the-lma}), and
it actually satisfies the detailed balance.  In this case, one
introduces a scalar function based on the steady state $\vx^{ss}$:
\begin{equation}
        A[\vx] =\sum_{j=1}^N
             \left[x_j(t)\ln\left(\frac{x_j(t)}{x_j^{ss}}\right)-x_j(t)
               +x^{ss}_j\right].
\label{macro-A}
\end{equation}
Using this function, it can be shown (see below)
that if a kinetic system has such an
{\em equilibrium steady state}, it is unique.
The equilibrium $\vx^{ss}$ in (\ref{macro-A})
can then be expressed in terms of intrinsic
properties of the chemical species, e.g., the $\mu$'s.
In the chemical kinetics literature,  the $A$[\vx] first
appeared as Shear's Liapunov function for kinetics
with detailed balance \cite{shear-jtb,higgins-1968}.
Horn and Jackson called it {\em pseudo-Helmholtz function}
for complex balanced but not detail balanced systems
\cite{horn-jackson}.

Second, if a steady state $\vx^{ss}$ is detail balanced, then $\sigma^{(hk)}\big[\vx\big]\equiv 0$ $\forall\vx$.
Otherwise, $\sigma^{(hk)}\big[\vx^{ss}\big]>0$ for any steady
state, stable or unstable. Then
there must be a concentration region, near $\vx^{ss}$, in which  $\sigma^{(hk)}\big[\vx\big]>0$.  Therefore,
$\sigma^{(hk)}[\vx]$ can only be negative, if ever,
when $\vx$ is {\em far from any steady state}.

	Third, $\sigma^{(fd)}\big[\vx\big]$ is
the time derivative of $A[\vx(t)]$ given in (\ref{macro-A}), when
$\vx(t)$ follows the rate equation (\ref{the-lma}):
\begin{eqnarray}
	 \frac{\rd A[\vx]}{\rd t}
      &=& \sum_{j=1}^N\frac{\rd x_j(t)}{\rd t}
                    \ln\left(\frac{x_j(t)}{x_j^{ss}}\right)
\nonumber\\
	        &=& \sum_{j=1}^N
        \left[ \sum_{\ell=1}^M  \Big(\kappa_{\ell j}-\nu_{\ell j}\Big)
                     \left( k_{+\ell}\prod_{i=1}^N
                  x_i^{\nu_{\ell i}} -k_{-\ell}\prod_{i=1}^N
                  x_i^{\kappa_{\ell i}} \right) \right]
                    \ln\left(\frac{x_j}{x_j^{ss}}\right)
\nonumber\\
	        &=& \sum_{\ell=1}^M
                     \left( k_{+\ell}\prod_{i=1}^N
                  x_i^{\nu_{\ell i}} -k_{-\ell}\prod_{i=1}^N
                  x_i^{\kappa_{\ell i}} \right)
                    \sum_{j=1}^N\ln\left(\frac{x_j}{x_j^{ss}}\right)^{\kappa_{\ell j}-\nu_{\ell j}}
\nonumber\\
	        &=& \sum_{\ell=1}^M
                     \Big(J_{+\ell} -J_{-\ell}\Big)
                   \ln \left(\frac{J_{-\ell}J_{+\ell}^{ss}}{J_{+\ell}J_{-\ell}(\vx^{ss})} \right)  \ = \ -\sigma^{(fd)}[\vx].
\end{eqnarray}
In other words,
\begin{equation}
      \sigma^{(fd)}[\vx] = -\sum_{i=1}^N
                \left(\frac{\partial A[\vx]}{\partial x_i}\right) F_i(\vx),
\label{eqn011}
\end{equation}
in which $\frac{\rd x_i}{\rd t}=F_i(\vx)=\sum_{\ell=1}^M
               \Big(\kappa_{\ell i}-\nu_{\ell i}\Big)
           \Big(J_{+\ell}(\vx)-J_{-\ell}(\vx)\Big)$, i.e. the right hand side of (\ref{the-lma}).  Noticing that
$A[\vx]$ attains its global minimum 0 at $\vx^{ss}$,
Eq. \ref{eqn011}
dictates $\nabla_{\vx}\sigma^{(fd)}[\vx^{ss}]=\bf{0}$.
To determine whether $\sigma^{(fd)}[\vx^{ss}]$ is a
maximum, minimum, or saddle point, we compute the
Hessian matrix $\mathcal{H}$:
\begin{eqnarray}
     \mathcal{H}_{ij}\big[\vx^{ss}\big]
  &\equiv&
    \frac{\partial^2 \sigma^{(fd)}[\vx^{ss}]}{\partial x_i\partial x_j}
\nonumber\\
	&=&
-\sum_{k=1}^N\left[
                \left(\frac{\partial^2 A[\vx^{ss}]}{\partial x_k\partial x_i}\right)
             \frac{\partial F_k(\vx^{ss})}{\partial x_j}
            + \left(\frac{\partial^2A[\vx^{ss}]}{\partial x_k\partial x_j}\right)
        \frac{\partial F_k(\vx^{ss})}{\partial x_i}\right]
\nonumber\\
   &=& -\frac{1}{x^{ss}_i}
                \frac{\partial F_i(\vx^{ss})}{\partial x_j}-\frac{1}{x^{ss}_j}\frac{\partial F_j(\vx^{ss})}{\partial x_i}.
\end{eqnarray}
Note matrix $\Gamma$, $\gamma_{ij}=\frac{\partial F_i(\vx^{ss})}{\partial x_j}$, defines the linear stability of $\vx^{ss}$.
If we denote $\Theta=\text{diag}\{(x_i^{ss})^{-1}\}$, then
\begin{equation}
 \mathcal{H}=-\big(\Theta\Gamma+\Gamma^{T}\Theta\big).
\label{keizerseq}
\end{equation}
There is a precise relationship between the Jacobian matrix and
Hessian matrix near an $\vx^{ss}$.

In one-dimensional case, $\mathcal{H}$ and $\Gamma$ always
have opposite signs, since a steady state is positive.
In high-dimensional case, however, even if
all the eigenvalues of $\Gamma$ are negative, which
implies the steady state $\vx^{ss}$ is stable, it is still
possible for a symmetric $\mathcal{H}$ to
have negative eigenvalues, resulting in negative $\sigma^{(fd)}[\vx]$ near a stable fixed point. A simple example of such is
\[
   \Gamma = \left(\begin{array}{cc}
           \ \ 1 &  \  \  2 \\ -3  &  -4
          \end{array}\right), \
      \Theta = \left(\begin{array}{cc}
      5 & 0 \\ 0 & 1 \end{array}\right),  \
          \mathcal{H} =  \left(\begin{array}{cc}
           \ \  -10 & -7 \\  -7  &  8
          \end{array}\right),
\]
in which matrix $\Gamma$ has eigenvalues $-1$ and
$-2$, but $\mathcal{H}$ has a negative eigenvalue
$-1-\sqrt{130}$.


{\bf\em  Concentrations of species with linear constraints.}
In chemical kinetics, the concentrations of many different
chemical species are often constrained by the
stoichiometric matrix.  In fact, matrix
$\mathcal{S}$, with elements
$s_{j\ell}=(\kappa_{\ell j}-\nu_{\ell j})$,
often has a high-dimensional left null space with vector
$(q_1,\cdots,q_N)$:
\begin{equation}
       \sum_{j=1}^N q_js_{j\ell}
      = \sum_{j=1}^N(\kappa_{\ell j}-\nu_{\ell j})
q_j=0,   \  \  \forall\ell.
\label{consrel}
\end{equation}
Each linearly independent null vector represents a conservation
of a certain chemical group in the entire
chemical reaction system:
\begin{equation}
   \frac{\rd}{\rd t}\sum_{j=1}^N q_jx_j(t) =
        \sum_{j=1}^N q_j\left(\frac{\rd x_j(t)}{\rd t}\right)
          = 0.
\label{equation15}
\end{equation}
If the left null space of $\mathcal{S}$ is $d$ dimensional,
then there are only $(N-d)$ independent
differential equations in the system (\ref{the-lma}).
The Jacobian matrix $\Gamma$ near a fixed point
$\vx^{ss}$ has a rank equal or lower than $(N-d)$.
As shown in Appendix \ref{app-B}, there
exists an $N\times(N-d)$ constant matrix $\mathcal{Z}$ with
rank $(N-d)$, the spanned space of whose column vectors are the same as the spanned space of the column vectors of $\mathcal{S}$.

In terms of the $\mathcal{Z}$ and an $(N-d)\times N$ constant matrix
$\mathcal{U}$, such that $\mathcal{U}\mathcal{Z}=I_{N-d}$.
Eq. \ref{the-lma} becomes $\frac{\rd}{\rd t}\vec{\delta}(t)=\vec{\mathcal{F}}\big(\vec{\delta}\big)$, in which
$\mathcal{Z}\vec{\delta}(t)=\vx(t)-\vx^{ss}$ and
$\vec{\mathcal{F}}\big(\vec{\delta}\big)=\mathcal{U}\vec{F}\big(\vx^{ss}+\mathcal{Z}
\vec{\delta}\big)$. $\vec{F}(\vx)=\{F_i(\vx)\}$ is the right hand side of Eq. (\ref{the-lma}).

Therefore, $\Gamma^{\delta}=\mathcal{U}\Gamma\mathcal{Z}$ is the linear matrix of the equation $\frac{\rd}{\rd t}\vec{\delta}(t)=\vec{\mathcal{F}}\big(\vec{\delta}\big)$ at $\vec{\delta}=0$, which determines the stability of the original steady state $\vx^{ss}$ constrained by the conservation relations. And the Hessian matrix $\mathcal{H}^{\delta}$ of $\sigma^{fd}$ with respect to the variable $\vec{\delta}$ without constrain becomes $\mathcal{H}^{\delta}=\mathcal{Z}^T\mathcal{H}\mathcal{Z}$.

Furthermore, since $\mathcal{Z}\mathcal{U}\mathcal{Z}=\mathcal{Z}$, $\mathcal{Z}\mathcal{U}$ is an identity mapping from the space spanned by the column vectors in $\mathcal{Z}$ to itself. Hence we have $\mathcal{S}=\mathcal{ZUS}$
and $\vec{F}=\mathcal{S}\vec{J}=\mathcal{ZUS}\vec{J}$, followed by $\mathcal{ZU}\Gamma=\Gamma$. Then
\begin{eqnarray}
    \mathcal{H}^{\delta} &=&
      \mathcal{Z}^T\mathcal{H}\mathcal{Z}  \ =  \
  -\mathcal{Z}^T\Big(\Theta\mathcal{ZU}\Gamma
       +\Gamma^{T}\mathcal{U}^T\mathcal{Z}^T\Theta
           \Big)\mathcal{Z}
\nonumber\\
	&=&  -\big(\mathcal{Z}^T\Theta\mathcal{Z}\big)\Gamma^{\delta}
       -\big(\Gamma^{\delta}\big)^T\big(\mathcal{Z}^T\Theta
           \mathcal{Z}\big),
\label{keizer2}
\end{eqnarray}
in which $(N-d)\times(N-d)$ metrix,
and symmetrix matrix
$\big(\mathcal{Z}^T\Theta\mathcal{Z}\big)$ is no
longer diagonal (compared with Eq. \ref{keizerseq}).


\subsection{Complex balanced kinetics and
non-negativity of $\sigma^{(hk)}$ and
$\sigma^{(fd)}$}
\label{sec:compba}

{\bf\em Macroscopic house-keeping heat.}
In stochastic thermodynamics, house-keeping heat
\cite{HS,ge-qian-pre10} is also known as adiabatic instantaneous
entropy production rate \cite{esposito2007,esposito-vandenbroeck}.
If the macroscopic reaction system (\ref{rxn}) is in a steady-state
$\vx^{ss}$, the $\ell^{th}$ reversible reaction has a chemical free energy dissipation per occurrence $\Delta\mu_{\ell}^{ss}=
k_BT\ln\big(J_{+\ell}(\vx^{ss})/J_{-\ell}(\vx^{ss})\big)$
\cite{ge-qian-pre13}.  Therefore,
in $k_BT$ unit and sum over all $M$ reversible reactions
we have
\begin{eqnarray}
    \sigma^{(hk)}\big[\vx\big]
	  &=&  \sum_{\ell=1}^M (J_{+\ell}(\vx)-J_{-\ell}(\vx))\frac{\Delta\mu_{\ell}^{ss}}{k_BT}
\nonumber\\
        &=&\sum_{\ell=1}^M \left[
             J_{+\ell}(\vx)\ln\left(
           \frac{J_{+\ell}(\vx^{ss})}{J_{-\ell}(\vx^{ss})}\right)
                +J_{-\ell}(\vx)\ln\left(
           \frac{J_{-\ell}(\vx^{ss})}{J_{+\ell}(\vx^{ss})}\right)
           \right]
\nonumber\\
   &\ge& \sum_{\ell=1}^M \left[ J_{+\ell}(\vx)\left(1-
           \frac{J_{-\ell}(\vx^{ss})}{J_{+\ell}(\vx^{ss})}\right)
                +J_{-\ell}(\vx)\left(1-
           \frac{J_{+\ell}(\vx^{ss})}{J_{-\ell}(\vx^{ss})}\right)
           \right]
\nonumber\\
       &=& \sum_{\ell=1}^M
            \Big(J_{+\ell}\big(\vx^{ss}\big)-
           J_{-\ell}\big(\vx^{ss}\big) \Big)  \left( \frac{ J_{+\ell}(\vx)}{J_{+\ell}\big(\vx^{ss}\big)}-
                 \frac{ J_{-\ell}(\vx) }{ J_{-\ell}\big(\vx^{ss}\big)}
                  \right).
\nonumber\\
\label{Qhk}
\end{eqnarray}
For an equilibrium steady state, detailed balance implies
$J_{+\ell}(\vx^{eq})-J_{-\ell}(\vx^{eq})=0$ for all $\ell$.
Therefore the rhs of (\ref{Qhk}) is zero for detailed balanced system.

{\bf\em Generalized free energy dissipation.}
Free energy dissipation \cite{esposito2007,ge-qian-pre10} is also known
as non-adiabatic instantaneous entropy production
rate \cite{esposito-vandenbroeck}, which is actually
the negative time-derivative of a generalized free energy
given in  (\ref{macro-A}).  This $A[\vx]$
figured prominently in Horn and Jackson's theory \cite{horn-jackson}.
Without the assumption of detailed balance,
\begin{eqnarray}
	 \sigma^{(fd)}[\vx]
	        &=& -\sum_{\ell=1}^M
                     \Big(J_{+\ell}(\vx) -J_{-\ell}(\vx)\Big)
                   \ln \left(\frac{J_{-\ell}(\vx)J_{+\ell}(\vx^{ss})}{J_{+\ell}(\vx)J_{-\ell}(\vx^{ss})} \right)
\nonumber\\
	        &=& -\sum_{\ell=1}^M \left[ J_{+\ell}(\vx)
                   \ln \left(\frac{J_{-\ell}(\vx)J_{+\ell}(\vx^{ss})}{J_{+\ell}(\vx)J_{-\ell}(\vx^{ss})} \right)  + J_{-\ell}(\vx)  \ln \left(\frac{J_{+\ell}(\vx)J_{-\ell}(\vx^{ss})}{J_{-\ell}(\vx)J_{+\ell}(\vx^{ss})} \right) \right]
\nonumber\\
	&\ge& -\sum_{\ell=1}^M \left[ J_{+\ell}
                 \left(\frac{J_{-\ell}(t)J_{+\ell}(\vx^{ss})}{J_{+\ell}(t)J_{-\ell}(\vx^{ss})}-1 \right)  + J_{-\ell}  \left(\frac{J_{+\ell}(t)J_{-\ell}(\vx^{ss})}{J_{-\ell}(t)J_{+\ell}(\vx^{ss})}-1 \right) \right]
\nonumber\\
	&=& \sum_{\ell=1}^M
                 \Big(J_{+\ell}\big(\vx^{ss}\big)-J_{-\ell}\big(\vx^{ss}\big)\Big)\left(\frac{J_{+\ell}(\vx)}{J_{+\ell}\big(\vx^{ss}\big)}-\frac{J_{-\ell}(\vx)}{J_{-\ell}\big(\vx^{ss}\big)}\right).
\end{eqnarray}
This is exactly the same rhs of (\ref{Qhk}).
Therefore, $\sigma^{(hk)}$ and $\sigma^{(fd)}$ are both
no-less than
\begin{eqnarray}
 && \sum_{\ell=1}^M
                 \Big(J_{+\ell}\big(\vx^{ss}\big)-J_{-\ell}\big(\vx^{ss}\big)\Big)\left(\frac{J_{+\ell}(\vx)}{J_{+\ell}\big(\vx^{ss}\big)}-\frac{J_{-\ell}(\vx)}{J_{-\ell}\big(\vx^{ss}\big)}\right)
\nonumber\\
	&=&  \sum_{\ell=1}^M
        \Big(J_{+\ell}\big(\vx^{ss}\big)-J_{-\ell}\big(\vx^{ss}\big)\Big)
    \left(  \prod_{i=1}^N \left(\frac{x_i}{x_i^{ss}}\right)^{\nu_{\ell i}}  -\prod_{i=1}^N \left(\frac{x_i}{x_i^{ss}}\right)^{\kappa_{\ell i}} \right).
\label{eq009}
\end{eqnarray}

{\bf\em Complex balanced chemical reaction
networks.}
A chemical reaction system is ``complex balanced'' if
and only if the rhs of (\ref{eq009}) is zero for any $\vx=(x_1,x_2,\cdots,x_N)$
\cite{horn-jackson,horn-1972,feinberg-1972}.
This is because any unique multi-type-nomial term
\[
      \prod_{i=1}^N
       \left(\frac{x_i}{x_i^{ss}}\right)^{\xi_i}
\]
represents a particular ``complex''
$(\xi_1X_1+\xi_2X_2+\cdots+\xi_NX_N)$.
Therefore, a complex
balanced steady state has all the influx to the complex
precisely balanced by the outflux of that complex:
\begin{equation}
        \left\{ \sum_{\ell=1}^M
     \Big(\delta_{\footnotesize\vkappa_{\ell},\vxi}-
             \delta_{\footnotesize\vnu_{\ell},\vxi} \Big)
        \Big( J_{+\ell}\big(\vx^{ss}\big)-J_{-\ell}\big(\vx^{ss}\big)\Big)
        \right\}
     \prod_{i=1}^N \left(\frac{x_i}{x_i^{ss}}\right)^{\xi_i}
          = 0.
\end{equation}
Detailed balance is a special case in which the
$J^{ss}_{+\ell}=J^{ss}_{-\ell}$ for every $\ell$.
Detailed balance is a
kinetic concept. Complex balance, however, has
a topological implication for a reaction network
\cite{feinberg-1972,feinberg-91}.

{\bf\em Lyapunov function of complex balanced kinetics.}
Since $\frac{\rd}{\rd t}A\big[\{\vx\}\big]=-\sigma^{(fd)}[\vx]\le 0$ for
a reaction network with complex balance, and
$A\big[\{\vx\}\big]\ge 0$,
it is a Lyapunov function for the mass-action
kinetics.  The convexity of $A[\{\vx\}]$ is easy
to establish:
$\partial^2A/\partial x_i\partial x_j=x_i^{-1}\delta_{ij}$.
Therefore, one concludes that the steady state $\vx^{ss}$
of a complex balanced reaction kinetics is unique.   This is
a well-known result and the proof was given in \cite{horn-jackson}.  The existence
of this Lyapunov function $A[\{\vx\}]$ for kinetic systems
with complex balance has prompted Horn and
Jackson's description of a ``quasithermodynamics''.

	In fact, an equally significant result is the following
statement:
\begin{quote}
{\em For reaction system with non-complex balanced kinetics,
if the macroscopic free energy dissipation $\sigma^{(fd)}[\vx]$ is
non-negative for all $\vx$, then the kinetics has a unique
steady state.  Equivalently, if a kinetic system is multi-stable, then
$\sigma^{(fd)}[\vx]$ is negative for some $\vx$, where
$\sigma^{(hk)}[\vx]>\sigma^{(tot)}[\vx]$.
}
\end{quote}

\section{Macroscopic limit of mesoscopic stochastic
thermodynamics}

\subsection{Kinetic description according to Delbr\"uck and Gillespie's model}

	The macroscopic chemical thermodynamics presented above does not
reference to anything with probability.  But the
very notion of Gibbs' chemical potential has a deep root in
it.  Chemical reactions at the individual molecule level are
stochastic \cite{wemoerner}.
A mathematically more accurate description of the chemical
kinetics in system (\ref{rxn}) is the stochastic theory of
Chemical Master Equation (CME) first appeared in the work of
Leontovich \cite{leontovich-35} and Delbr\"uck \cite{delbruck},
whose fluctuating trajectories can be exactly computed using the
stochastic simulation method widely known as Gillespie
algorithm \cite{gillespie}.  Note these two descriptions are
not two different theories, rather they are the two aspects of
a same Markov process, just as the diffusion equation
and the Langevin-equation descriptions of a same Brownian
motion.  More importantly, this probabilistic description and
Waage-Guldberg's law of mass action are also two parts of a
same dynamic theory: The latter is the limit of the former if
fluctuations are sufficiently small, when the volume of
the reaction system, $V$, is large \cite{beard-qian-book}.
In fact, the key quantity in Delbr\"uck-Gillespie's description
of mesoscopic chemical kinetics is the rate of a particular
reaction, called propensity function.  For the $\ell^{th}$
forward and backward reactions, they are
\begin{subequations}
\begin{eqnarray}
    u_{\ell}(\vn) &=& k_{+\ell} V\prod_{j=1}^n
            \left( \frac{n_j!}{(n_j-\nu_{\ell j})!V^{\nu_{\ell j}}}\right),
\\
	w_{\ell}(\vn) &=& k_{-\ell} V\prod_{j=1}^n
            \left( \frac{n_j!}{(n_j-\nu_{\ell j})!
               V^{\kappa_{\ell j}}}\right).
\end{eqnarray}
One sees that in the macroscopic limit
$V\rightarrow\infty$, $V^{-1}u_{\ell}(V\vx) =J_{+\ell}(\vx)$
and $V^{-1}w_{\ell}(V\vx) =J_{-\ell}(\vx)$, where $\vx=\vn/V$.
\end{subequations}

The stochastic trajectory can be expressed in terms of the
random-time-changed Poisson representation:
\begin{eqnarray}
    &&  n_j(t) \  = \  n_j(0) +
\\
  &&
\sum_{\ell=1}^M
          \Big(\kappa_{\ell j}-\nu_{\ell j}\Big)
             \left\{        Y_{+\ell}\left(
                    \int_0^t u_\ell\Big(\vn(s)\Big) \rd s \right)
             -  Y_{-\ell}\left(
                    \int_0^t w_\ell\Big(\vn(s)\Big) \rd s \right) \right\},
\nonumber
\end{eqnarray}
where $Y_{\pm\ell}(t)$ are $2\ell$ independent, standard
Poisson processes:
\begin{equation}
  \Pr\big\{ Y (t) = n\big\} = \frac{t^n}{n!}e^{-t},  \
            Y(0)=0.
\end{equation}
The CME for the mesosocpic kinetics is
\begin{eqnarray}
  \frac{\rd p(\vn,t)}{\rd t} &=& \sum_{\ell=1}^M
       \Big[   p(\vn-\vkappa_{\ell}+\vnu_{\ell})u_{\ell}(\vn-\vkappa_{\ell}+\vnu_{\ell})
\nonumber\\
      && - p(\vn)\Big( u_{\ell}(\vn)+w_{\ell}(\vn) \Big)
                 + p(\vn+\vkappa_{\ell}-\vnu_{\ell})
                    w_{\ell}(\vn+\vkappa_{\ell}-\vnu_{\ell})  \Big].
\label{the-cme}
\end{eqnarray}
\begin{subequations}
	The mesoscopic stochastic thermodynamics provides the
following set of equations \cite{ge-qian-pre10}:
\begin{eqnarray}
    e_p\big[p(\vn,t)\big] &=& \sum_{\ell=1}^M \sum_{\vn}
                  \Big(p(\vn+\vkappa_{\ell})w_{\ell}(\vn+\vkappa_{\ell})- p(\vn+\vnu_{\ell}) u_{\ell}(\vn+\vnu_{\ell}) \Big)
\nonumber\\
	&& \times\ln\left(\frac{p(\vn+\vkappa_{\ell})w_{\ell}(\vn+\vkappa_{\ell})}{p(\vn+\vnu_{\ell}) u_{\ell}(\vn+\vnu_{\ell}) }\right)
\\
  &=& f_d\big[p(\vn,t)\big]  + Q_{hk}\big[p(\vn,t)\big] ,
\\
	f_d\big[p(\vn,t)\big] &=& \sum_{\ell=1}^M \sum_{\vn}
                  \Big(p(\vn+\vkappa_{\ell})w_{\ell}(\vn+\vkappa_{\ell})- p(\vn+\vnu_{\ell}) u_{\ell}(\vn+\vnu_{\ell}) \Big)
\nonumber\\
	&& \times\ln\left(\frac{p(\vn+\vkappa_{\ell}) p^{ss}(\vn+\vnu_{\ell}) }{p(\vn+\vnu_{\ell}) p^{ss}(\vn+\vkappa_{\ell})}\right) \ = \ -\frac{\rd F^{(\text{meso})}}{\rd t},
\\
	F^{(\text{meso})}\big[p(\vn,t)\big] &=&  \sum_{\vn} p(\vn,t)\ln\left(
             \frac{p(\vn,t)}{p^{ss}(\vn)}\right),
\\
	Q_{hk}\big[p(\vn,t)\big] &=&  \sum_{\ell=1}^M \sum_{\vn}
                  \Big(p(\vn+\vkappa_{\ell})w_{\ell}(\vn+\vkappa_{\ell})- p(\vn+\vnu_{\ell}) u_{\ell}(\vn+\vnu_{\ell}) \Big)
\nonumber\\
	&& \times\ln\left(\frac{p^{ss}(\vn+\vkappa_{\ell})w_{\ell}(\vn+\vkappa_{\ell})}{p^{ss}(\vn+\vnu_{\ell}) u_{\ell}(\vn+\vnu_{\ell}) }\right).
\end{eqnarray}
All three $e_p\big[p(\vn,t)\big] $, $f_d\big[p(\vn,t)\big] $ and $Q_{hk}\big[p(\vn,t)\big] $ $\ge 0$ \cite{ge-qian-pre10,esposito-vandenbroeck}.
\end{subequations}

\subsection{Macroscopic limits}

	Denote $\vx(t)$ as the solution of the corresponding deterministic model (Eq. \ref{the-lma}). In the macroscopic limit when $\vn,V\rightarrow\infty$, $\vx=\vn/V$, one has the asymptotic expressions \cite{Kurtz1978}
\begin{equation}
                    p(V\vx,t) \simeq \frac{1}{V}\delta(\vx-\vx(t)), \  \textrm{ and }   \
                    p_V^{ss}(V\vx) \simeq e^{-V\varphi^{ss}(\vx)},
\end{equation}
where $\delta(\vx-\vx(t))$ is the $\delta$ measure concentrating at $\vx(t)$, $p_V^{ss}(\cdot)$ is the stationary distribution of the chemical master equation (\ref{the-cme}) and $\inf_{\vx}\varphi^{ss}(\vx)=0$.
As a part of the theory of large deviations,
the mathematical existence of $\varphi^{ss}(\vx)$ has
been extensively discussed \cite{Shwartz1995,Kurtz1978}.  See recent \cite{anderson-2015}
and references cited within.  Therefore,
\begin{eqnarray}
    e_p &\simeq& V\sum_{\ell=1}^M \int \rd\vx
                   \delta(\vx-\vx(t)) \Big(J_{-\ell}(\vx)-J_{+\ell}(\vx)\Big)\times 
         \ln\left(\frac{J_{-\ell}(\vx)}{J_{+\ell}(\vx) }\right)\nonumber\\
          &\rightarrow& V\sigma^{(tot)}\big[\vx(t)\big],
\end{eqnarray}
where the {\em density} of macroscopic chemical
entropy production rate
\begin{equation}
	\sigma^{(tot)}[\vx] = \sum_{\ell=1}^M \Big(J_{-\ell}(\vx)-J_{+\ell}(\vx) \Big)
            \ln\left(\frac{J_{-\ell}(\vx)}{
               J_{+\ell}(\vx) }\right).
\end{equation}
This is Eq. \ref{sigmatot}. 

	On the other hand,
\begin{eqnarray}
	f_d &=& \sum_{\ell=1}^M \sum_{\vn}
                  \Big(p(\vn+\vkappa_{\ell})w_{\ell}(\vn+\vkappa_{\ell})- p(\vn+\vnu_{\ell}) u_{\ell}(\vn+\vnu_{\ell}) \Big)
\nonumber\\
	&& \times\ln\left(\frac{p(\vn+\vkappa_{\ell}) p^{ss}(\vn+\vnu_{\ell}) }{p(\vn+\vnu_{\ell}) p^{ss}(\vn+\vkappa_{\ell})}\right)
\nonumber\\
	&\simeq& V\sum_{\ell=1}^M \int\rd\vx \delta(\vx-\vx(t))
                  \Big(J_{-\ell}(\vx)-J_{+\ell}(\vx) \Big) \ln \frac{p_V^{ss}(\vx+\vnu_{\ell})}{p_V^{ss}(\vx+\vkappa_{\ell})}
\nonumber\\
 &\simeq& V\sum_{\ell=1}^M \int\rd\vx \delta(\vx-\vx(t))
                  \Big(J_{-\ell}(\vx)-J_{+\ell}(\vx) \Big)
         \big(\vnu_{\ell}-\vkappa_{\ell}\big)\cdot
       \nabla_{\vx} \ln p_V^{ss}(\vx)
\nonumber\\
   &\rightarrow& V \sum_{\ell=1}^M
                  \Big(J_{-\ell}(\vx(t))-J_{+\ell}(\vx(t)) \Big)
         \big(\vkappa_{\ell}-\vnu_{\ell}\big)\cdot
       \nabla_{\vx} \varphi^{ss}(\vx(t)).
\label{oo}
\end{eqnarray}

Comparing with Eq. (\ref{eqn011}), we thus have the following statement:
\begin{quote}
{\em $\sigma^{(fd)}[\vx(t)]$ and the macroscopic
limit of $V^{-1}f_d\big[p(V\vx,t)\big]$ are equal if and
only if the $\varphi^{ss}(\vx)$ and the $A[\vx]$
in (\ref{macro-A}) differ by a conserved quantity
of (\ref{the-lma}).}
\end{quote}
That is,
$\varphi^{ss}(\vx)=A[\vx]+C[\vx]$ where
$\vec{F}(\vx)\cdot\nabla_{\vx}C[\vx]=0$ \cite{polettini-esposito-1}.
Once the limit of $V^{-1}f_d(t)\neq \sigma^{(fd)}[\vx(t)]$, it implies that the
limit of $V^{-1}Q_{hk}(t)$ will not be $\sigma^{(hk)}[\vx(t)]$.

	Last, but not the least,
\begin{eqnarray}
Q_{hk}&=&e_p-f_d\nonumber\\
&\simeq&V \sum_{\ell=1}^M \Big(J_{-\ell}(\vx)-J_{+\ell}(\vx) \Big)
            \ln\left(\frac{J_{-\ell}(\vx)}{
               J_{+\ell}(\vx) }\right)\nonumber\\
               &&-V\sum_{\ell=1}^M
                  \Big(J_{-\ell}(\vx)-J_{+\ell}(\vx) \Big)
         \big(\vkappa_{\ell}-\vnu_{\ell}\big)\cdot
       \nabla_{\vx} \varphi^{ss}(\vx)\nonumber\\
       &=& V \sum_{\ell=1}^M \Big(J_{-\ell}(\vx)-J_{+\ell}(\vx) \Big)
            \left[\ln\left(\frac{J_{-\ell}(\vx)}{
               J_{+\ell}(\vx) }\right)-\big(\vkappa_{\ell}-\vnu_{\ell}\big)\cdot
       \nabla_{\vx} \varphi^{ss}(\vx)\right]\nonumber\\
       &=& V\sum_{\ell=1}^M \Big(J_{-\ell}(\vx)-J_{+\ell}(\vx) \Big)
            \ln\left(\frac{J_{-\ell}(\vx)}{
               J_{+\ell}(\vx) }e^{\big(\vnu_{\ell}-\vkappa_{\ell}\big)\cdot
       \nabla_{\vx} \varphi^{ss}(\vx)}\right).
\label{eq026}
\end{eqnarray}
Therefore, the macrocopic limit of $V^{-1}Q_{hk}$ contains
the
\begin{equation}
      \ln\left(\frac{J_{-\ell}(\vx)}{
               J_{+\ell}(\vx) }e^{\big(\vnu_{\ell}-\vkappa_{\ell}\big)\cdot
       \nabla_{\vx} \varphi^{ss}(\vx)}\right),
\label{eq0027}
\end{equation}
which in turn is dependent upon the unknown function $\varphi^{ss}(\vx)$;
$\varphi^{ss}(\vx)$ is an emergent, global quantity itself.
In contrast, $\sigma^{(hk)}$ depends upon only local rate laws
\begin{equation}
            \ln\left(\frac{J_{-\ell}(\vx^{ss})}{
               J_{+\ell}(\vx^{ss}) }\right),
\label{eq0028}
\end{equation}
which is the expression (\ref{eq0027}) evaluated at
$\vx=\vx^{ss}$, where $\nabla_{\vx}\varphi^{ss}\big[\vx^{ss}\big]=0$.

	For kinetic systems with complex balanced,
Anderson et.al. have recently shown that $A[\vx]=\varphi^{ss}(\vx)$\cite{anderson-2015}.
Therefore, Eqs. \ref{eq0027} and \ref{eq0028} are indeed the
same:
\begin{eqnarray*}
    && \ln\left(\frac{J_{-\ell}(\vx)}{
               J_{+\ell}(\vx) }e^{\big(\vnu_{\ell}-\vkappa_{\ell}\big)\cdot
       \nabla_{\vx} A[\vx]}\right)
\nonumber\\
	&=&  \ln\left(\frac{J_{-\ell}(\vx)}{
               J_{+\ell}(\vx) } \right)
          + \sum_{j=1}^N
       \ln\left(\frac{x_j}{x_j^{ss}}\right)^{\nu_{\ell j}-\kappa_{\ell j}}
 \  = \ \ln\left(\frac{J^{ss}_{-\ell}(\vx)}{
               J^{ss}_{+\ell}(\vx) } \right).
\end{eqnarray*}
The non-local (\ref{eq0027}) is reduced to the local (\ref{eq0028})
in this case.

Eqs. \ref{eq0027} and \ref{eq0028} are the same if and only if $A[\vx]=\varphi^{ss}(\vx)$. However, since the macroscopic limit of $V^{-1}\big(f_d(t)+Q_{hk}\big)$ is the same as $\sigma^{(fd)}+\sigma^{(hk)}$, the macroscopic limit of $V^{-1}Q_{hk}$ is the same as $\sigma^{(hk)}$ if and only if the $\varphi^{ss}(\vx)$ and the $A[\vx]$
in (\ref{macro-A}) differ by a conserved quantity of (\ref{the-lma}).

\subsection{Nonequilibrium free energy and its time derivative}

In the theory of mesoscopic, stochastic thermodynamics,
the generalized nonequilibrium free energy \cite{ge-qian-pre10}
\begin{eqnarray}
 F^{(\text{meso})} &=&  \sum_{\vn} p(\vn,t)\ln\left(
             \frac{p(\vn,t)}{p^{ss}(\vn)}\right).
\end{eqnarray}
In the macroscopic limit, its density therefore is
\begin{eqnarray}
		&& \frac{1}{V}\sum_{\vn} p(\vn,t)\ln\left(\frac{p(\vn, t)}{
                   p^{ss}(\vn)}\right)
              \  \simeq  \   \frac{1}{V}\int f_V(\vx,t)
                \ln\left(\frac{f_V(\vx,t)}{f_V^{ss}(\vx)}
                   \right) \rd\vx
\nonumber\\
	&\simeq& \frac{1}{V}\int f_V(\vx,t)\ln\left( f_V(\vx,t)e^{V\varphi^{ss}(\vx)}
                  \int e^{-V\varphi^{ss}(\vz)} \rd\vz \right) \rd\vx
\nonumber\\
	&=& \int f_V(\vx,t) \varphi^{ss}(\vx) \rd\vx
          +\frac{1}{V}\int f_V(\vx,t)\ln f_V(\vx,t)\rd\vx
\nonumber\\
       &&   +\frac{1}{V}\ln\left( \int e^{-V\varphi^{ss}(\vz)} \rd\vz\right)
\label{aaa}\\
	&\rightarrow&   \varphi^{ss}\big(\vx(t)\big).
\label{bbb}
\end{eqnarray}
One can recognize the first two terms in the rhs of Eq. \ref{aaa}
as the instantaneous mean internal energy and entropy, and
the last term as the logarithm of a partition function, with
$V$ playing the role of the $\beta$.\footnote{A separation
of the first, instantaneous mean energy, and the
last stationary free energy naturally arises in the mathematical
limit: For finite $V$, $ -V^{-1}\ln f^{ss}_V(\vx)\simeq$
$\varphi^{ss}(\vx)$ $+$ log-partition function,
where $\inf_{\vx} \varphi^{ss}(\vx)=0$. The partition
function, therefore, provides an appropriate energy
reference point for a macroscopic system.  This is the spirit of
renormalization; its fundamental insight resides in the
notion of conditional probability.  H.Q. thank Dr. Panagiotis Stinis
for an illuminating discussion on the theory of
renormalization.}    Therefore,
\begin{quote}
{\em The macroscopic limit of the
mesoscopic, stochastic thermodynamic free energy
is the chemical potential of mean force for
the macroscopic dynamics}
\cite{santillan-qian,esposito-13}.
\end{quote}

For relatively simple kinetics with complex balance,
the $A[\vx]$ in (\ref{macro-A}) and the $\varphi^{ss}(\vx)$
in (\ref{bbb})
being the same \cite{anderson-2015} signifies
a complete consistency between the kinetics and
thermodynamics.  For such systems, the $\varphi^{ss}(\vx)$
has a generic, simple form, i.e. $A[\vx]$.  For complex kinetics,
however, the $\varphi^{ss}(\vx)$ is not known {\em a
priori}.  An accurate computation of the
$\varphi^{ss}(\vx)$, as an emergent quantity,  has to be
computationally demanding.

\subsection{Keizer's macroscopic nonequilibrium
thermodynamics}

For complex balanced kinetics,
it can be shown (see Appendix \ref{app-C})
that the matrix relation in Eq. \ref{keizerseq}
is in fact consistent with J. Keizer's macroscopic, local
nonequilibrium thermodynamics \cite{keizer} which
states:
$2D=-\big(\Gamma\Xi+\Xi\Gamma^T\big)$,
in which $\Xi^{-1}=\Theta$, the curvature of $A[\vx]$
at $\vx^{ss}$, and
\begin{equation}
         D(\vx^{ss}) =\frac{1}{2}\sum_{\ell=1}^M \big(\nu_{\ell i}-\kappa_{\ell i}\big)
           \big(\nu_{\ell j}-\kappa_{\ell j}\big)\Big(
                J_{+\ell}(\vx^{ss})+J_{-\ell}(\vx^{ss})\Big).
\end{equation}

Eq. \ref{keizer2}, however, indicates that for complex balanced
kinetic systems, the correlations between the
concentrations fluctuations near the steady state are a simple
consequence of the conservation relations in (\ref{consrel}).
For example,
$A+B\ \rightleftharpoonsfill{22pt}\ C$
has $\mathcal{Z}=(1,1,-1)^T$, and
$\mathcal{Z}^T\Theta\mathcal{Z}=(x_A^{ss})^{-1}+
(x_B^{ss})^{-1}+(x_C^{ss})^{-1}$.
Such correlations are simple; they are fundamentally different
from the correlations that arise in complex dynamics such as chemical
oscillations \cite{qian-pnas-02}.

\section{Discussion}

	It is well-known that entropy is not the appropriate thermodynamic
potential for isothermal systems; free energy is: Helmholtz's for
constant volume and Gibbs' for constant pressure.   Therefore, it is
not surprising that while our equation
$\rd A/\rd t=-\sigma^{(tot)}+\sigma^{(hk)}$ is similar to
the celebrated entropy balance equation
$\rd S/\rd t = \sigma^{(tot)}+\eta^{(ex)}$, where $\eta^{(ex)}$ is the
rate of entropy exchange of the system with its surrounding
\cite{degroot-mazur-book,qkkb},
the new equation is a more appropriate description of the nonequilibrium,
thermodynamics at a constant temperature.   As a consequence,
positive $\sigma^{(hk)}$ and $\sigma^{(tot)}$
can be interpreted as the source and the sink of the chemical
free energy of a reaction system with complex balance.
For non-driven chemical  reaction system, $\sigma^{(hk)}=0$,
$\rd A/\rd t = -\sigma^{(tot)}$, and $\eta^{(ex)}$ is the
time derivative of the total internal energy.

	Macroscopic ``laws'' are emergent properties of
complex dynamics at a level below: This is a
profound insight from studies in many-body physics
\cite{pwanderson,jjh,laughlin}.  In the present work, we observe
precisely how this idea works in terms of the $\varphi^{ss}(\vx)$
as a statistical ``summary'' of the mesoscopic, long time
dynamics; yet it serves as a ``law of force'' for the macroscopic
behavior.  For relatively simple systems, the $\varphi^{ss}(\vx)$
can be known {\em a priori}, with a {\em generic, robust} form,
as the Gibbs free energy for the case of detailed balance
non-driven chemical systems, and the $A[\vx]$ for
complex balanced systems.  Both systems have been known for
a long time to be uni-stable.
For more complex kinetics, however, it is impossible to know
$\varphi^{ss}(\vx)$ {\em a priori}.  It is a true emegent quantity
that requires to be discovered.

{\bf Acknowledgment}
We thank R. Rao and M. Esposito for sharing their
interesting paper \cite{rao-esposito} with us.
H. Ge is supported by NSFC (No. 21373021), and 863 program (No. 2015AA020406).

\appendix

\section{Gibbs and Helmholtz free energies, concentration scales}
\setcounter{equation}{0}
\renewcommand\theequation{A\arabic{equation}}

	These materials are not new, we summarize them
here since they are not easy to find in the literature. Since the Helmholtz free energy
$A(T,V,\{n_i\})=G(T,p,\{n_i\})-pV$, $(\partial A/\partial n_i)_{T,V}=(\partial G/\partial n_i)_{T,p}=\mu_i$, in which $G(T,p,\{n_i\})$ is the Gibbs free energy.  Therefore, no matter a system is maintained at a constant volume or constant pressure, the equilibrium condition derived from minimizing the corresponding free energy is the same for a reaction,
i.e., $\Delta\mu_{\ell}=0$.  Furthermore, for dilute solution, the difference $pV$ is nearly not changing with time \cite{Fermi},
therefore in practice it is immaterial whether one considers the equilibrium condition at constant volume or at constant pressure.

For constant pressure, the chemical potential $\mu_i$
should be expressed in mole fraction.  For a dilute solution consist of $n_0$ moles of solvent and $n_i$ moles of the $i^{th}$ solute ($n_i\ll n_0$), the total Gibbs free energy is defined as \cite{Fermi}
$G=\sum_{i=0}^Nn_i\mu_i$, in which $\mu_i=\mu_i^o+k_BT\log (n_i/n_t)=\partial G/\partial n_i$, where $n_t=n_0+n_1+\cdots+n_N$.
$\mu_i$ is the per molecule chemical potential of the $i^{th}$ component.

{\bf{\boldmath{$(a)$}} Constant pressure.} We first show for the case of constant pressure, where components are expressed in terms of
mole fraction:
$G=\sum_{i=0}^N n_i\mu_i$ in which component $0$ is the solvent.
For dilute solution $n_0\simeq n_t$.
Therefore,  $\mu_0=\hat{\mu}_0^o+k_BT\ln(n_0/n_t)$
$\simeq\hat{\mu}^o_0-$$k_BT\sum_{i=1}^N(n_i/n_0)$.  Therefore,
the Gibbs free energy can
be approximated as \cite{Fermi}
\begin{equation}
 G = n_0\hat{\mu}^o_0+\sum_{i=1}^N n_i\left[
                   \hat{\mu}_i^o-k_BT
          +k_BT\ln\left(\frac{n_i}{n_0}\right) \right],
\end{equation}
where $\mu_i=\hat{\mu}_i^o+k_BT\ln(n_i/n_t)$
$\simeq\hat{\mu}_i^o+k_BT\ln(n_i/n_0)$ for each $i\geq 1$.
Without loss of generality, we can set $\hat{\mu}_0^o=0$, hence $G=\sum_{i=1}^N n_i\big(\mu_i-k_BT\big)$. Define $\mu_i^o=\hat{\mu}_i^o-k_BT\ln(n_0/V)$ for each $i\geq 1$, then $\mu_i=\mu_i^o+k_BT\ln(n_i/V)$. Hence per unit volume and
in units of $k_BT$: $G=\sum_{i=1}^N x_i\mu_i-\sum_{i=1}^Nx_i$. This is Eq. \ref{G4solution}.

We now show Eq. \ref{lyapunov11}. For chemical reaction:
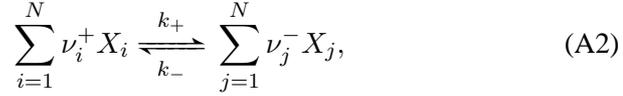
\begin{equation}
      \sum_{i=1}^N \nu^+_iX_i
      \  \underset{k_-}{\overset{k_+}{\rightleftharpoonsfill{26pt}}}   \
          \sum_{j=1}^N \nu^-_jX_j,
\label{rxn-in-appendix}
\end{equation}
through minimizing $G$ we can have \cite{Fermi}:
$\sum_{i=1}^N\nu^+_i\mu_i$
$=\sum_{j=1}^N\nu^-_j\mu_j$, i.e.
$$\sum_{i=1}^N \nu^+_i\left[\mu_i^o+k_BT\ln\left(\frac{n_i}{V}\right)_{eq}\right]
=\sum_{j=1}^N \nu^-_j\left[\mu_j^o+k_BT\ln\left(\frac{n_j}{V}\right)_{eq}\right].$$
Hence the equilibrium constant
$$\frac{k_-}{k_+}= \prod_{i=1}^N \left(\frac{n_i}{V}\right)_{eq}^{\nu^+_i-\nu^-_i}=\exp\left[\frac{1}{k_BT}\left(\sum_{j=1}^N \nu^-_j\mu_j^o-\sum_{i=1}^N \nu^+_i\mu_i^o\right)\right],$$
results in
$$k_BT\ln\left(\frac{J_-}{J_+}\right)=\sum_{j=1}^N \Big(\nu^-_j-\nu^+_j\Big)\mu_i,$$
which guarantees the decreasing of Gibbs free energy with time.


{\bf{\boldmath{$(b)$}} Constant volume.}  We now rewrite $G=\sum_{i=0}^N n_i\mu_i$, in which $\mu_i=\bar{\mu}_i^o+k_BT\ln(n_i/V)$ is the per molecule
chemical potential of the $i^{th}$ component, in molarity, including the solvent as the $0^{th}$ component.  For chemical
equilibrium one minimizes the Helmholtz free energy $A$ rather
than the Gibbs free energy $G$.

$$A=G-pV=\sum_{i=0}^N n_i(\mu_i-p_iV)=\sum_{i=0}^N n_i\tilde{\mu_i},$$
in which $p_i$ is the partial pressure for one molecule of the $i^{th}$ component, and $\tilde{\mu_i}=\tilde{\mu}_i^o+k_BT\ln(n_i/V),\tilde{\mu}_i^o=\bar{\mu}_i^o-p_iV$.

	Again for the chemical reaction in (\ref{rxn-in-appendix})
through minimizing $A$ we can have in equilibrium
\[
           \sum_{i=1}^N \nu^+_i\big(\tilde{\mu}_i+k_BT\big)=\sum_{j=1}^s\nu^-_j\big(\tilde{\mu}_j+k_BT\big).
\]
Hence the equilibrium constant
$$\frac{k_-}{k_+}
=\exp\left[\frac{1}{k_BT} \sum_{j=1}^N \Big(\nu^-_j-\nu^+_j\Big)\tilde{\mu}_j^o+\sum_{j=1}^N \Big(\nu^-_j-\nu^+_i\Big)\right].$$
We can here define $\check{\mu}_i=\tilde{\mu}_i+k_BT$ for each $i\geq 1$, then $A=\sum_{i=1}^Nn_i\check{\mu}_i-k_BT\sum_{i=1}^Nn_i$.   Hence
$$\frac{k_-}{k_+}=\exp\left(\frac{1}{k_BT}
     \sum_{j=1}^N \Big(\nu^-_j-\nu^+_j\Big) \check{\mu}_j^o\right),$$
in which $\check{\mu}_i^o=\tilde{\mu}_i^o+k_BT$.
The form here is the same as that in the previous case now.

We can eliminate $n_0\tilde{\mu}_0=-n_0k_BT$ from $A$, then following this condition, we can have the Helmholz free energy
decreasing with time, not the Gibbs free energy.

One can see in this case Eq. \ref{eq11} is just the Helmhotz free energy. In dilute solution, $A\simeq G + \text{const}$.

\section{Kinetics with conserved quantities}
\label{app-B}
\setcounter{equation}{0}
\renewcommand\theequation{B\arabic{equation}}

Denote $\mathcal{L}\subseteq \mathbb{R}^N_{+}$ as the left null space of the stoichiometric matrix $\mathcal{S}=\{s_{i\ell}=\kappa_{\ell i}-\nu_{\ell i}\}_{N\times M}$.
Any vector in $\mathcal{L}$ represents a conservation law of the chemical reaction system, i.e. for each $\vec{q}=(q_1,q_2,\cdots,q_N)$ satisfies Eq. \ref{equation15}: $\vec{q}\ \mathcal{S}\vec{J}=0$,
in which $\vec{J}=(J_1,J_2,\cdots,J_M)^{T}$,
$J_{\ell}=J_{+\ell}(\vx)-J_{-\ell}(\vx)$.

Suppose the dimension of $\mathcal{L}$ is $d$, hence the dimension of the span of column vectors (also of the matrix $\mathcal{S}$) in $\mathcal{S}$ which is orthogonal to $\mathcal{L}$ is $N-d$. Given the $d$ conservation laws according to the basis of the space $\mathcal{L}$, the deterministic and stochastic dynamics of the chemical reaction system is constrained in this subspace.

Suppose the $(N-d)$ linearly independent column vectors of $\mathcal{S}$ as $\vz_i=(z_{i1},z_{i2},\cdots,z_{iN})^{T}$, $i=1,\cdots,N-d$. Hence given a steady state values $\vx^{ss}=(x_1^{ss},\cdots,x_N^{ss})^{T}$, we rewrite the dynamics of the chemical reaction system using the new variables $\delta=(\delta_1,\cdots,\delta_{N-d})^{T}$ which satisfies
\[ \vx(t)=\vx^{ss}+\sum_{j=1}^{N-d} \delta_j(t)\vz_j, \ \text{ or }\
   x_i(t)=x^{ss}_i+\sum_{j=1}^{N-d} \delta_j(t)z_{ji}.
\]

Let $N\times (N-d)$ matrix $\mathcal{Z}$
taking $\vz_i$ as the column vectors, hence $\vx(t)-\vx^{ss}=\mathcal{Z}\vec{\delta}(t)$. Since the dimension of $\mathcal{Z}$ is $(N-d)$, if there is a vector $\vec{\delta}$ satisfying $\vx-\vx^{ss}=\mathcal{Z}\vec{\delta}$, then it is unique. Therefore, for any matrix $\mathcal{U}=\{u_{ij}\}_{(N-d)\times N}$ satisfying $\mathcal{U}\mathcal{Z}=I_{N-d}$, we can have
$\vec{\delta}=\mathcal{U}(\vx-\vx^{ss})$.

The deterministic dynamics of $\vec{\delta}(t)$ becomes
\begin{equation}
\frac{\rd \vec{\delta}(t)}{\rd t}=\mathcal{U}\frac{\rd\vx(t)}{\rd t}=\mathcal{U}\mathcal{S}\vec{\mathcal{J}}\big(\vec{\delta}(t)\big),
\label{ind-ode}
\end{equation}
in which $\vec{\mathcal{J}}\big(\vec{\delta}\big)=\vec{J}\big(\vx^{ss}+\mathcal{Z}\vec{\delta}\big)$.
The system Eq. \ref{ind-ode} is a set of $(N-d)$ independent ordinary
differential equations, while the system in (\ref{the-lma}) usually
is not.  Unfortunately, the intrinsic chemical kinetic structure is lost in
the transformation of (\ref{the-lma}) to (\ref{ind-ode}).

Let $\vec{\mathcal{F}}\big(\vec{\delta}\big)=
\mathcal{U}\mathcal{S}\vec{\mathcal{J}}\big(\vec{\delta}\big)
=\mathcal{U}\vec{F}\big(\vx^{ss}+\mathcal{Z}\vec{\delta}\big)$, in which $N$ dimensional vector
$\vec{F}(\vx)=\mathcal{S}\vec{J}(\vx)$ is the right hand side of the original Eq. \ref{the-lma}. Hence
a new Jacobian matrix, $\Gamma^\delta=\mathcal{U}\Gamma\mathcal{Z}$, has
elements
\begin{equation}
     \gamma_{ij}=\frac{\partial F^\delta_i(0)}{\partial \delta j}=\sum_{\ell=1}^{N-d}u_{i\ell}\sum_{k=1}^N\frac{\partial F_\ell(\vx^{ss})}{\partial x_k}z_{kj},
\end{equation}
$1\le i,j\le N-d$.

Therefore, suppose $\vec{\eta}$ is an eigenvector of $\Gamma$ with eigenvalue $\lambda$. If $\lambda\neq 0$, $\vec{\eta}$ is in the space $\mathcal{L}^{+}$; hence there exists a corresponding vector $\vec{\eta}^\delta$ satisfying $\mathcal{Z}\vec{\eta}^\delta=\vec{\eta}$,
followed by $\Gamma^\delta\vec{\eta}^\delta=\lambda\vec{\eta}^\delta$. So the dimensions of eigenspaces of nonzero eigenvalues for
$N\times N$ Jacobian matrix $\Gamma$ and $(N-d)\times (N-d)$
matrix
$\Gamma^\delta$ are the same. This implies that the sufficient and necessary conditions for the steady state $\vx^{ss}$ to be stable are the real parts of all eigenvalues of $\Gamma^\delta$ are negative, i.e. the eigenspace of zero eigenvalue for $\Gamma$ has dimension $d$, and all the nonzero eigenvalues have negative real parts.  Note that the remaining
$\Gamma^{\delta}$ is still possible to have zero eigenvalue(s) at
$\vec{\delta}=0$ due to dynamics.  This yields a center
manifold.

Next we will consider  $\sigma^{(fd)}\big[\vx\big]$.

Denote $\sigma_\delta^{(fd)}(\vec{\delta})=\sigma^{(fd)}\big[\vx^{ss}+\mathcal{Z}\vec{\delta}\big]$. Then
\begin{equation}
     \frac{\partial \sigma_\delta^{(fd)}}{\partial \delta_i}\big(\vec{\delta}\big)
  =\sum_{j=1}^N\frac{\partial \sigma^{(fd)}\big[\vx^{ss}+\mathcal{Z}\vec{\delta}\big]}{\partial x_j}z_{ji}.
\end{equation}
Therefore, we know that at $\vx=\vx^{ss}$, i.e. $\vec{\delta}=\bf{0}$, \[
    \frac{\partial \sigma_\delta^{(fd)}}{\partial \delta_i}({\bf{0}})
 =\frac{\partial \sigma^{(fd)}}{\partial \vx_i}\big[\vx^{ss}\big]=0,
\]
for each $i$.  Furthermore, let us compute $\frac{\partial^2 \sigma_\delta^{(fd)}}{\partial \delta_i\partial \delta_j}({\bf{0}})$.
Since
\begin{equation}
   \frac{\partial \sigma_\delta^{(fd)}}{\partial \delta_i}\big(\vec{\delta}\big)=\sum_{k=1}^N\frac{\partial \sigma^{(fd)}\big[\vx^{ss}+\mathcal{Z}\vec{\delta}\big]}{\partial x_k}z_{ki},
\end{equation}
we have
\begin{eqnarray}
\frac{\partial^2 \sigma_\delta^{(fd)}}{\partial \delta_i\partial \delta_j}\big(\vec{\delta}\big)
 &=&\sum_{k=1}^Nz_{ki}\frac{\partial\left[\frac{\partial \sigma^{(fd)}\left[\vx^{ss}+\mathcal{Z}\vec{\delta}\right]}{\partial x_k}\right]}{\partial \delta_j}
\nonumber\\
&=&\sum_{k=1}^Nz_{ki}\sum_{m=1}^N\frac{\partial^2 \sigma^{(fd)}}{\partial x_k\partial x_m}\left[\vx^{ss}+\mathcal{Z}\vec{\delta}\right]z_{mj}.
\end{eqnarray}
Hence define matrix $\mathcal{H}^\delta=\{H_{ij}^\delta\}$, in which $H_{ij}^\delta=\frac{\partial^2 \sigma_\delta^{(fd)}}{\partial \delta_i\partial \delta_j}({\bf{0}})$, we have
$\mathcal{H}^\delta=\mathcal{Z}^T\mathcal{H}\mathcal{Z}$.

If $\mathcal{H^\delta}$ has negative eigenvalue, $\sigma^{(fd)}(\vx)$ can be negative at certain $\vx$ around $\vx^{ss}$.  Conversely,
if $\sigma^{(fd)}(\vx)$ can be negative for some $\vx$ in arbitrarily small neighborhood of $\vx^{ss}$, then $\mathcal{H^\delta}$ must have negative eigenvalue.

\section{Keizer's theory with complex balanced
kinetics}
\label{app-C}
\setcounter{equation}{0}
\renewcommand\theequation{C\arabic{equation}}

According to Keizer's theory \cite{keizer}, at a steady state $\vx^{ss}$ with complex balance,
 $2D=-(\Gamma\Xi+\Xi\Gamma^T)$, in which $\Theta=\left\{\frac{\partial A[\vx^{ss}]}{\partial x_i\partial x_j}\right\}$,
\begin{eqnarray}
  \Xi_{ij} &=& \left(\Theta^{-1}\right)_{ij} \ = \ x^{ss}_i\delta_{ij},
\\
   \Gamma_{ij} &=&  \frac{\partial F_i(\vx^{ss})}{\partial x_j}
\nonumber\\
	&=&  \frac{1}{x_j}\sum_{\ell=1}^M \Big(\kappa_{\ell i}-\nu_{\ell i} \Big)
       \Big(\nu_{\ell j}J_{+\ell}(\vx^{ss})
       -\kappa_{\ell j}J_{-\ell}(\vx^{ss})\Big),
\\
  D_{ij} &=& \frac{1}{2}\sum_{\ell=1}^M \big(\nu_{\ell i}-\kappa_{\ell i}\big)
           \big(\nu_{\ell j}-\kappa_{\ell j}\big)\Big(
                J_{+\ell}(\vx^{ss})+J_{-\ell}(\vx^{ss})\Big).
\label{c4}
\end{eqnarray}
We note that $\Xi=\Theta^{-1}$ in Eq. \ref{keizerseq}, and
\begin{eqnarray}
   && -\sum_{k=1}^N \Big(\Gamma_{ik} x^{ss}_k\delta_{kj}+x^{ss}_i\delta_{ik}\Gamma_{jk}\Big) \ = \   -\big( \Gamma_{ij}x^{ss}_j+\Gamma_{ji}x^{ss}_i \big)
\nonumber\\
 &=& \sum_{\ell=1}^M \Big(
       \big(\nu_{\ell i} -\kappa_{\ell i}\big)\nu_{\ell j}
          +\big(\nu_{\ell j} -\kappa_{\ell j}\big)\nu_{\ell i}   \Big)
          J_{+\ell}(\vx^{ss})
\nonumber\\
    &&   -\Big( \big(\kappa_{\ell i}-\nu_{\ell i} \big)\kappa_{\ell j}
            + \big(\kappa_{\ell j}-\nu_{\ell j} \big)\kappa_{\ell i}\Big)
         J_{-\ell}(\vx^{ss}).
\label{b3}
\end{eqnarray}

Now for each complex $(\xi_1X_1+\cdots+\xi_NX_N)$, represented
by $\vxi$, a steady state $\vx^{ss}$ being complex balanced
means
\begin{equation}
  \sum_{\ell=1}^M  \Big(
      \psi(\vkappa_{\ell})\delta_{\vkappa_{\ell},\vxi}-
       \psi(\vnu_{\ell})\delta_{\vnu_{\ell},\vxi}\Big)\Big(J_{+\ell}\big(\vx^{ss}\big)
                  -J_{-\ell}\big(\vx^{ss}\big)\Big)
       = 0,
\end{equation}
for any functions $\psi$.  If we choose
$\psi(\vxi)=\xi_1\xi_j$, then
\begin{equation}
\sum_{\ell=1}^M \Big(
       \kappa_{\ell i}\kappa_{\ell j}-\nu_{\ell j}\nu_{\ell i}
      \Big) \Big(J_{+\ell}(\vx^{ss})-J_{-\ell}(\vx^{ss})\Big)
   =  0.
\label{b5}
\end{equation}

Combining (\ref{b3}) and (\ref{b5}), we have
\begin{eqnarray}
 && \sum_{\ell=1}^M \Big(
       \big(\nu_{\ell i} -\kappa_{\ell i}\big)\nu_{\ell j}
          +\big(\nu_{\ell j} -\kappa_{\ell j}\big)\nu_{\ell i}   \Big)
          J_{+\ell}(\vx^{ss})
  -\Big( \big(\kappa_{\ell i}-\nu_{\ell i} \big)\kappa_{\ell j}
\nonumber\\
   && +\big(\kappa_{\ell j}-\nu_{\ell j} \big)\kappa_{\ell i}\Big)
         J_{-\ell}(\vx^{ss})
   + \Big(\kappa_{\ell i}\kappa_{\ell j}-\nu_{\ell j}\nu_{\ell i}
      \Big) \Big(J_{+\ell}(\vx^{ss})-J_{-\ell}(\vx^{ss})\Big)
\nonumber\\
  &=& \sum_{\ell=1}^M  \big(\nu_{\ell i}-\kappa_{\ell i}\big)
           \big(\nu_{\ell j}-\kappa_{\ell j}\big)\Big(
                J_{+\ell}(\vx^{ss})+J_{-\ell}(\vx^{ss})\Big)
\nonumber\\
  &=& 2D_{ij}.
\end{eqnarray}

\end{document}